\documentclass[aps,prl,twocolumn,groupedaddress]{revtex4-1}

\usepackage{aas_macros}
\usepackage{graphicx}

\usepackage{color}

\begin{document}


\title{Updated observational constraints on quintessence dark energy models}


\author{Jean-Baptiste Durrive$^{1}$, Junpei Ooba$^{1}$, Kiyotomo Ichiki$^{1,2}$ and Naoshi Sugiyama$^{1,2,3}$}
\affiliation{$^{1}$ Department of Physics and Astrophysics, Nagoya University, Nagoya 464-8602, Japan\\
$^2$ Kobayashi-Maskawa Institute for the Origin of Particles and the Universe, Nagoya University, Nagoya 464-8602, Japan\\
$^3$ Kavli Institute for the Physics and Mathematics of the Universe (Kavli IPMU), The University of Tokyo, Chiba 277-8582, Japan
}%

\email[]{jean.baptiste.durrive@e.mbox.nagoya-u.ac.jp}

\date{\today}

\begin{abstract}
The recent GW170817 measurement favors the simplest dark energy models, such as a single scalar field. Quintessence models can be classified in two classes, freezing and thawing, depending on whether the equation of state decreases towards $-1$ or departs from it. In this paper we put observational constraints on the parameters governing the equations of state of tracking freezing, scaling freezing and thawing models using updated data, from the Planck 2015 release, joint light-curve analysis and baryonic acoustic oscillations. Because of the current tensions on the value of the Hubble parameter $H_0$, unlike previous authors, we let this parameter vary, which modifies significantly the results. Finally, we also derive constraints on neutrino masses in each of these scenarios.
\end{abstract}


\maketitle


\section{I. Introduction}

The recent acceleration of the expansion of the Universe remains a mystery. This puzzle gave rise to the concept of dark energy (DE), an additional constituent in the Universe of unknown nature. From observations, the only property we are sure of is that, if it exists, it must have an equation of state very close to $w = -1$. A most natural way of going beyond the cosmological constant description of DE (for which $w = -1$ throughout cosmic history) is to consider a canonical minimally coupled scalar field, dubbed quintessence. It is all the more important to focus on this simple description that the recent observation of the binary neutron star merger GW170817 by the LIGO-VIRGO Collaboration and its associated electromagnetic counterpart by Fermi \cite{AbbottEtAl17} implies that gravitational waves travel at a speed extremely close to that of light, which favors the simplest models for dark energy and modified gravity, as discussed in \cite{CreminelliVernizzi17} and \cite{EzquiagaZumalacarregui17} for instance.

Many quintessence models have been proposed already; see, for example, \cite{Tsujikawa13} for a review. A convenient and efficient way of analyzing the variety of cosmological dynamics of quintessence is a dynamical system approach, cf.~\cite{AmendolaTsujikawa10}. Instead of having to consider each model separately, we may classify them into essentially two classes, namely freezing and thawing models, depending on whether the equation of state decreases towards $-1$ or departs from it as the scale factor grows \citep{CaldwellLinder05}. In addition, freezing models can themselves be subdivided into so-called tracking and scaling models, depending on the details of the dynamics. In this paper, following previous works, we constrain the parameters arising in three analytic expressions for the equation of state, corresponding to, respectively, tracking freezing, scaling freezing and thawing dynamics, so that our analysis covers most quintessence potentials. Figure~\ref{fig:EoS} shows the typical profiles of the equations of state we considered.

Now, by construction, quintessence necessarily has an equation of state that cannot go below $-1$. However in this paper we consider both the case of quintessence, indeed putting the prior $w \geq -1$ in our analysis, but we also extend the study to phantom DE, i.e. fluids having equations of state below $-1$ \citep[see e.g.][for studies discussing the viability of such cases]{ChibaEtAl09,CreminelliEtAl09,DuttaEtAl09}.

An important feature of the present study is that, compared to previous authors, we let the crucial Hubble parameter $H_0$ vary. Doing so is all the more important nowadays that there are currently tensions between the supernovae \cite{RiessEtAl16} and Planck results \cite{Planck16} in determining its precise value. Most recently an independent measure was brought in by the LIGO-VIRGO collaboration, which confirms that it must be around 70, but that measurement is not yet competitive with the two aforementioned ones \cite{LIGOVIRGOH017}.

This paper is organized as follows. In Sec.~II,
after very briefly reviewing the equations governing quintessence dark energy, we first present each of the three parametrizations (tracking, scaling and thawing) that we considered, and then expose the method and data we used for the analysis. In Sec.~III
 we show and discuss the observational constraints we obtain on the parameters intervening in these parametrizations, by considering them one after the other. Finally, we present observational constraints on the sum of neutrino masses by performing again the previous analysis but without assuming only massless neutrinos.

\section{II. Setup}
\label{section:SetUp}

\subsection{A. Quintessence}

Quintessence is a canonical scalar field $\phi$ minimally coupled to gravity. Considering, in addition, nonrelativistic matter, the action reads
\begin{equation}
S = \int d^4x \sqrt{-g} \left[\frac{M_\mathrm{pl}^2}{2} R - \frac{1}{2} g^{\mu \nu} \partial_\mu \phi \partial_\nu \phi - V(\phi)\right] + S_\mathrm{m},
\end{equation}
where $g$ is the determinant of the metric $g_{\mu \nu}$, $M_\mathrm{pl}$ is the Planck mass, $R$ is the Ricci scalar, $V(\phi)$ is the potential of the scalar field, and $S_\mathrm{m}$ is the action for the matter component. Considering a flat Friedmann-Lemat\^{i}re-Robertson-Walker background with $a(t)$ denoting the scale factor, the equations of motion are
\begin{equation}
\left\{
\begin{array}{l}
3 H^2 M_\mathrm{pl}^2 = \rho_\phi + \rho_\mathrm{m}\\
\ddot{\phi} + 3 H \dot{\phi} + V_{,\phi} = 0\\
\dot{\rho}_m + 3 H \rho_\mathrm{m} = 0
\end{array}
\right.
\end{equation}
where dots are derivatives with respect to cosmic time, $H = \dot{a}/a$ is the Hubble parameter, $\rho_\phi$ is the energy density of dark energy, $\rho_\mathrm{m}$ that of the nonrelativistic matter, and $V_{,\phi}$ denotes a derivative with respect to $\phi$. The equation of state is defined as
\begin{equation}
w = \frac{P_\phi}{\rho_\phi}
\label{EoS}
\end{equation}
where the pressure and energy density of the field are, respectively, given by
\begin{equation}
\left\{
\begin{array}{l}
P_\phi = \frac{\dot{\phi}^2}{2} - V(\phi)\\
\rho_\phi = \frac{\dot{\phi}^2}{2} + V(\phi).
\end{array}
\right.
\label{Pphi_rhophi}
\end{equation}
Finally, the density parameter is defined as $\Omega_{\phi} = \rho_\phi / (3 H^2 M_\mathrm{pl}^2)$. It is clear from eqs. (\ref{EoS}) and (\ref{Pphi_rhophi}) that $w \geq -1$ for quintessence. We will refer to this condition as the `quintessence prior' in the analysis below. The precise dynamics of quintessence depends on the details of the potential $V(\phi)$. Let us now detail the three cases we will consider in this paper. They are representative cases, such that our analysis covers the various possible behaviors of quintessence in general.

\begin{figure}
\includegraphics[scale=.9]{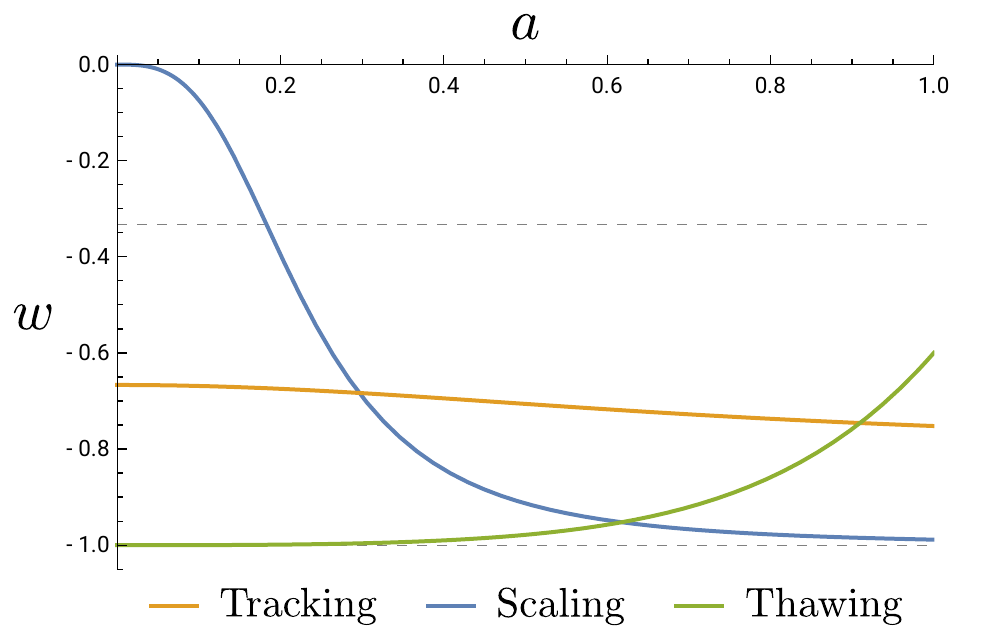}
\caption{Typical examples of the evolution $w(a)$ of the equation of state with the scale factor for the three types of dynamical dark energy models considered in this analysis: Tracking freezing models have an evolution given by (\ref{EoS_Tracking}), scaling freezing models by (\ref{EoS_Scaling}), and thawing models by (\ref{EoS_Thawing}). For reference, the upper horizontal dashed gray line corresponds $w=-1/3$, delimiting regions with and without cosmic acceleration, while the lower one corresponds to the phantom divide line $w=-1$, which is a lower limit in the case of quintessence. For illustration in this figure we chose $\Omega_{\phi 0}=0.7$, $p=1$ for the tracking case, $a_t = 0.23$ for the scaling case and ($K=2.9$,$w_0=-0.6$) for the thawing case.}
\label{fig:EoS}
\end{figure}

\subsubsection{Tracking freezing models}

In freezing models the field rolled down along the potential in the past, gradually slowing down as the system enters the acceleration phase. An important subclass of such models are so-called tracking models, such as in the case of a power-law potential
\begin{equation}
V(\phi) = M^{4+p} \phi^{-p}
\end{equation}
where $p>0$. This arises in the fermion condensate model as a dynamical supersymmetry breaking \cite{Binetruy99} for instance. For such types of potentials the condition $V V_{,\phi \phi}/V_{,\phi}^2>1$, called the tracking condition, is satisfied (see e.g. \cite{AmendolaTsujikawa10}) so that the field density eventually catches up that of the background fluid. The characteristic of such systems is that for a wide range of initial conditions, solutions converge to a common evolution, the tracker solution \cite{2002PhRvD..65f3502Y}. This feature is interesting in order to solve the coincidence problem. In the precise case of the power-law potential with nonrelativistic matter, one can show that (see e.g. \cite{Tsujikawa13})
\begin{equation}
w_{(0)} = - \frac{2}{p+2},
\label{Link_w0_p}
\end{equation}
so that observational constraints on $w_{(0)}$ actually translate into constraints on the exponent $p$. In addition, by perturbing the tracker equation, \cite{Chiba10} derived the equation of state for tracker fields to all orders in $\Omega_\phi$. In this work we considered that expression up to the third order, namely,
\begin{equation}
w(a) = w_{(0)} + \alpha_1 \Omega_\phi(a) + \alpha_2 \Omega_\phi(a)^2 + \alpha_3 \Omega_\phi(a)^3
\label{EoS_Tracking}
\end{equation}
where
\begin{equation}
\left\{
\begin{array}{l}
\alpha_1 =  \frac{(1-w_{(0)}^2) w_{(0)}}{1-2 w_{(0)} + 4 w_{(0)}^2}\\
\alpha_2 =  \frac{(1-w_{(0)}^2) w_{(0)}^2 (8 w_{(0)} - 1)}{(1-2 w_{(0)} + 4 w_{(0)}^2) (1-3 w_{(0)} + 12 w_{(0)}^2)}\\
\alpha_3 =  \frac{2 (1-w_{(0)}^2) w_{(0)}^3 (4 w_{(0)} - 1) (18 w_{(0)} + 1)}{(1-2 w_{(0)} + 4 w_{(0)}^2) (1-3 w_{(0)} + 12 w_{(0)}^2) (1-4 w_{(0)} + 24 w_{(0)}^2)}\\
\end{array}
\right.
\end{equation}
and
\begin{equation}
\Omega_\phi(a) = \frac{\Omega_{\phi 0} a^{-3 w_{(0)}}}{1-\Omega_{\phi 0} + \Omega_{\phi 0} a^{-3 w_{(0)}}}.
\end{equation}
However the second order approximation is already very good compared to the numerical solution as detailed in \cite{Chiba10}, and we checked that adding the third order term actually does not change the observational constraints significantly. Note that the parameter $w_{(0)}$ parametrizes the value of the equation of state during the matter-dominated era, and not the present day value, hence the different notation than $w_0$ used for the thawing case below. A typical example of this $w(a)$ evolution is plotted in Fig.~\ref{fig:EoS} for illustration.

\subsubsection{Scaling freezing models}

In this second type of freezing model \citep{CopelandEtAl98}, the equation of state scales as the background fluid, here matter. For a simple exponential potential, the expansion would never start accelerating. However for instance for a double exponential potential \citep{BarreiroEtAl00}
\begin{equation}
V(\phi) = V_1 e^{-\lambda_1 \phi / M_\mathrm{pl}} + V_2 e^{-\lambda_2 \phi / M_\mathrm{pl}}
\end{equation}
with $\lambda_1 \gg 1$ and $\lambda_2 \ll 1$, the potential is first dominated by the exponential with $\lambda_1$, so that the solution scales as matter in the early matter dominated epoch, but in a second stage, at late times, the other exponential in the potential dominates, so that the solution does enter a dark-energy dominated era, with cosmic acceleration as required. This transition occurs at a redshift which depends on the parameters $\lambda_1, \lambda_2, V_1$ and $V_2$. Unfortunately, for scaling freezing models there is currently no analytic formula for $w$ derived directly from the dynamics of the scalar field, like those used here for tracking and thawing cases. However, like in \cite{ChibaEtAl13}, we note that the equation of state in this case is in fact well approximated by the parametrization \cite{LinderHuterer05}
\begin{equation}
w(a) = -1 + \frac{1}{1 + (a/a_t)^{1/\tau}}
\label{EoS_Scaling}
\end{equation}
where $a_t$ denotes the scale factor of the onset of the transition and $\tau$ fixes the thickness of the transition effectively. In \cite{ChibaEtAl13} the authors found that $\tau=0.33$ is an appropriate choice for this analytic expression to fit the numerical solution very well, which we thus also make in this work. A typical example of this $w(a)$ evolution is plotted in Fig.~\ref{fig:EoS} for illustration.

\subsubsection{Thawing models}

In this class of models, the field, of mass $m_\phi$, was frozen by the Hubble friction $H \dot{\phi}$ in the past, such that $w = -1$, until $H$ drops below $m_\phi$ at a recent epoch, after which it begins to evolve and its equation of state departs from $-1$. A representative potential is the Hilltop potential~\citep{DuttaScherrer08}
\begin{equation}
V(\phi) = \Lambda^4 [1 + \cos(\phi / f)]
\label{Hilltop}
\end{equation}
relevant in pseudo-Nambu-Goldstone boson models \citep{FriemanEtAl95} of quintessence for example.

The scalar potential needs to be shallow enough for the field to evolve slowly along the potential, in a similar way as in inflationary cosmology. In \cite{Chiba09}, the author derived the slow-roll conditions for thawing quintessence, generalizing the work of \citep{DuttaScherrer08} who focused on the case of a hilltop potential only. By solving the equation of motion for the field $\phi$, Taylor expanding the potential around its initial value up to the quadratic order, in the limit where the equation of state is close to $-1$, the author derived the equation of state as a function of the scale factor, namely
\begin{equation}
\begin{array}{rl}
w(a) = & -1 + (1+w_0) a^{3(K-1)}\\
& \displaystyle \hspace{-0.5cm}
\times \left[\frac{(K-F)(F+1)^K + (K+F)(F-1)^K}{(K-F_0)(F_0+1)^K + (K+F_0)(F_0-1)^K}\right]^2
\end{array}
\label{EoS_Thawing}
\end{equation}
where
\begin{equation}
F(a) = \sqrt{1+(\Omega_{\phi 0}^{-1}-1) a^{-3}},
\end{equation}
with its present value
\begin{equation}
F_0 = F(a_0) = \Omega_{\phi 0}^{-1/2},
\end{equation}
and the constant K is given by
\begin{equation}
K = \sqrt{1 - \frac{4 M_\mathrm{pl}^2 V_{,\phi\phi}(\phi_i)}{3 V(\phi_i)}}.
\end{equation}
Physically, this constant $K$ is related to the mass of the field (second derivative of the potential with respect to $\phi$) at the initial stage. Following \cite{ChibaEtAl13} we note that this analytic expression of $w(a)$ agrees well with the numerical solution as long as $0.1 \lesssim K \lesssim 10$; therefore, in our analysis, we put this range as a prior on the parameter $K$. A typical example of the evolution $w(a)$ from (\ref{EoS_Thawing}) is plotted in Fig.~\ref{fig:EoS}.

The equation of state (\ref{EoS_Thawing}) contains three parameters: $w_0$, $K$ and $\Omega_{\phi 0}$. We stress that when confronting the models to data, by varying the parameter $K$ in the analysis we do not need to specify the shape of the potential as long as it is thawing, as detailed in \cite{Chiba09}. Therefore it is a fairly general analysis.

\subsection{B. Method of analysis}

We modify the publicly available Boltzmann code CLASS \cite{BlasEtAl11} to include our various equations of state. We then perform analyses using the MonteCarlo code MontePython \cite{AudrenEtAl12}. In all our runs, we make the cosmological parameters vary (the density parameters for baryons $\Omega_\mathrm{b}$ and for cold dark matter $\Omega_\mathrm{cdm}$, the peak scale parameter $100 \ \theta_\mathrm{s}$, the amplitude $A_\mathrm{s}$ and spectral index $n_\mathrm{s}$
of primordial curvature fluctuations and the reionization optical depth $\tau_\mathrm{reio}$), plus the nuisance parameters relevant to each experiment considered, plus the parameters relevant for each model under study (namely $w_{(0)}$ for the tracking case, $a_t$ for the scaling case and $w_0$ and $K$ for the thawing case), plus the sum of neutrino masses $M_\nu$ in the last part of this paper in which we performed the runs again, relaxing the assumption of massless neutrinos made otherwise. For all of those parameters we consider flat prior. In this paper we then present our results after marginalizing over all unmentioned parameters, and discuss them with and without the quintessence prior $w \geq -1$. Note that to realize the late-time cosmic acceleration, $w$ below $-1/3$ is also required, though we do not put it as a prior, since the data should rule out models without such feature.

The data and corresponding likelihoods we used are those from the Planck 2015 release \cite{Planck15} (temperature and polarization TT, TE, EE, and also the lensing which we present separately in our results), Supernovae (SDSS-II/SNLS3 Joint Light-curve Analysis (JLA) \cite{BetouleEtAl14}) and Baryonic Acoustic Oscillations (BAO) measurements (Sloan Digital Sky Survey Main Galaxy Sample SDSS7 MGS \cite{RossEtAl15}, Six-degree-Field Galaxy Survey 6dFGS \cite{BeutlerEtAl11}, Baryon Oscillation Spectroscopic Survey BOSS LOWZ and BOSS CMASS \cite{AndersonEtAl14}).

\section{III. Results and discussion}
\label{Results}

In this section we present and discuss the observational constraints we obtain on the three parametrizations of quintessence mentioned in the previous section.

\subsection{A. Tracking freezing models}

\begin{figure}
  \includegraphics[width=1.\linewidth]{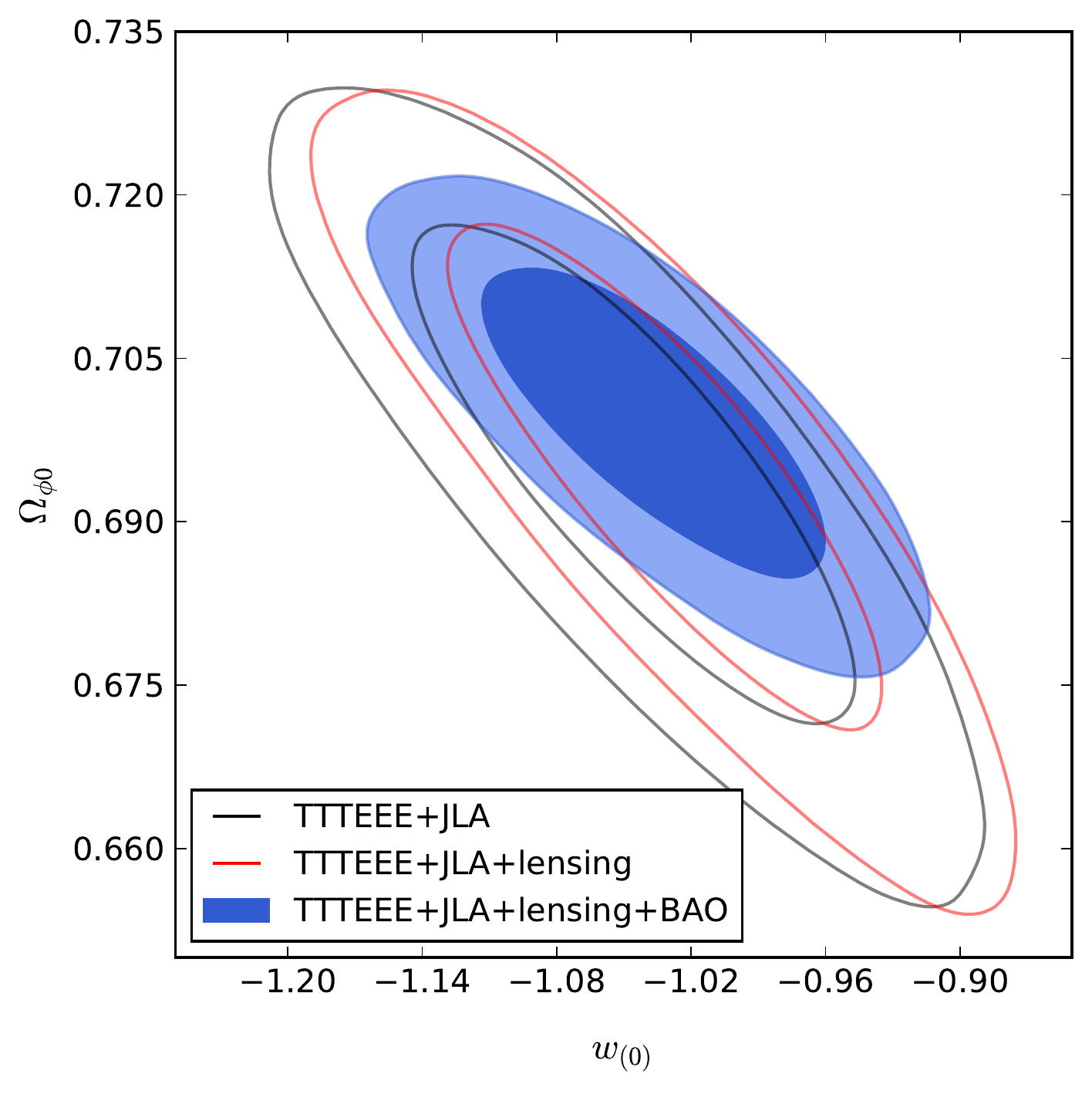}
  \caption{Observational constraints on models with tracking equation of state (\ref{EoS_Tracking}), in the $(w_{(0)},\Omega_{\phi 0})$ plane, having marginalized over all the other parameters. The red and black lines represent the constraints without BAO, with the inner lines corresponding to $1\sigma$ contours and the outer lines the $2\sigma$ ones. The filled contours in blue are the constraints including BAO measurements.}
  \label{fig:Contours_Track}
\end{figure}

\begin{figure}
  \includegraphics[width=1.05\linewidth]{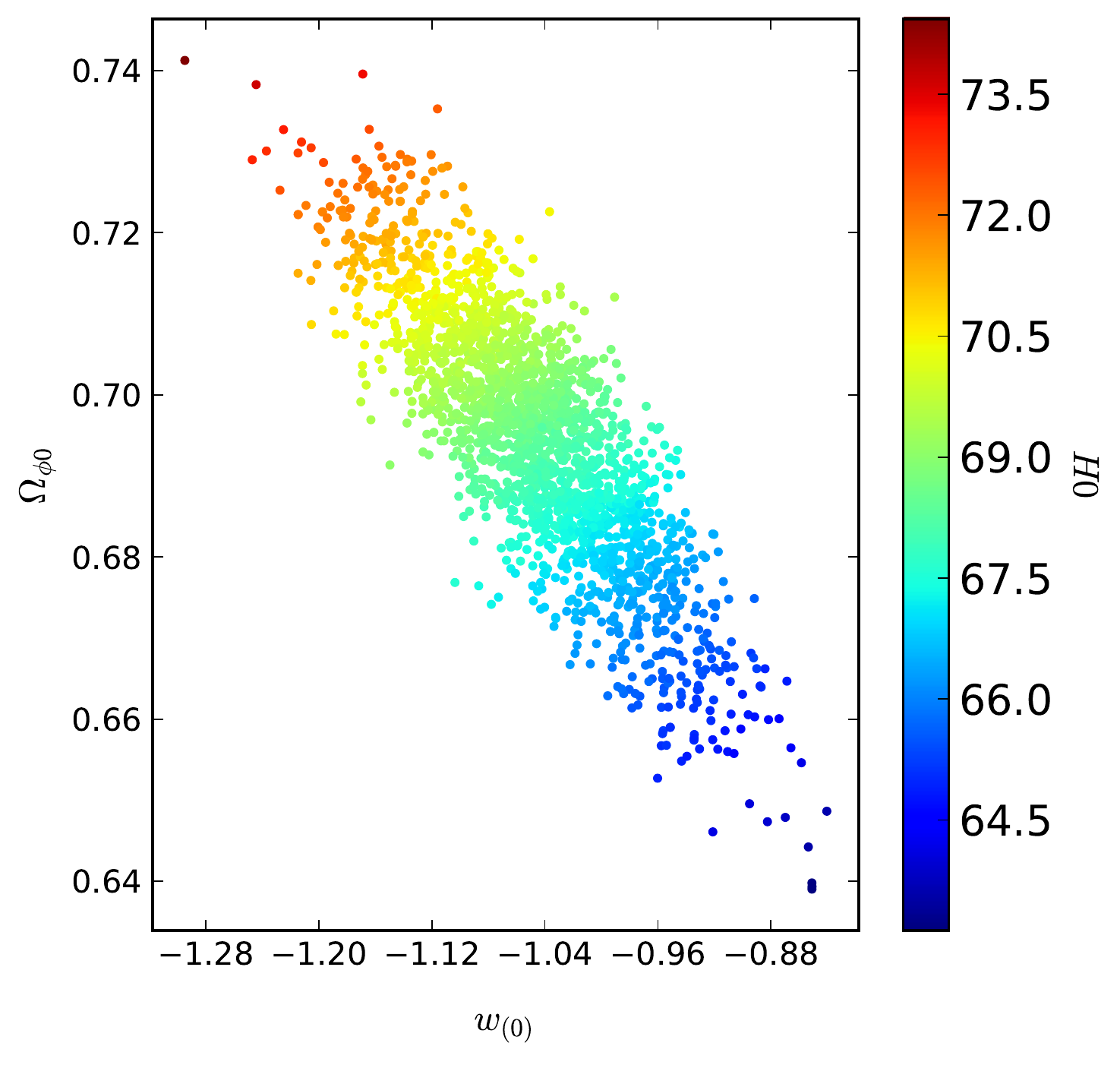}
  \caption{Confidence region from Fig.\ref{fig:Contours_Track} (the TTTEEE + JLA case) as a 3D plot, to explicit the values of parameter $H_0$. In all models studied here, it is a generic feature that the greater $H_0$, the more negative the equation of state. This plot also accounts for the fact that we obtain a direction of degeneracy that is perpendicular to that obtained in \cite{ChibaEtAl13}.}
  \label{fig:3dPlot}
\end{figure}

Figure \ref{fig:Contours_Track} shows the constraints we obtain in the tracking freezing case. It corresponds to the $1\sigma$ and $2\sigma$ confidence regions in the $(w_{(0)},\Omega_{\phi 0})$ plane, without the quintessence prior, in three cases, namely with the Planck polarization and supernovae data only, then adding the lensing information and finally adding the BAO measurements too. We observe that the BAO data improve significantly the constraints, while the lensing has a relatively small impact. After marginalizing over $w_{(0)}$ we get, without the quintessence prior,
\begin{equation}
0.680 < \Omega_{\phi 0} < 0.718 \ (95 \% \ \mathrm{C.L.})
\end{equation}
while with the prior,
\begin{equation}
0.675 < \Omega_{\phi 0} < 0.703 \ (95 \% \ \mathrm{C.L.}).
\end{equation}
Now on the contrary, marginalizing over $\Omega_{\phi 0}$, we get, without the quintessence prior,
\begin{equation}
-1.141 < w_{(0)} < -0.933 \ (95 \% \ \mathrm{C.L.})
\end{equation}
while with the prior,
\begin{equation}
-1 < w_{(0)} < -0.923 \ (95 \% \ \mathrm{C.L.}).
\end{equation}
Comparing our Fig.~\ref{fig:Contours_Track} with the equivalent figure in \cite{ChibaEtAl13} may be surprising, because in that paper the direction of degeneracy is perpendicular to ours. To account for that, we plot in Fig.~\ref{fig:3dPlot} a three-dimensional plot which includes the information on the value of $H_0$ (in this illustrative example the contours are given in the case of Planck polarization and JLA data only). The point is that in \cite{ChibaEtAl13} the authors performed their analysis with a fixed value of $H_0$, and it is clear from Fig.~\ref{fig:3dPlot} that fixing the value of that parameter results in a rotated direction of the ellipses. Actually, this figure is also interesting to notice that small values of $H_0$ corresponds to $w_{(0)}$ closer to zero. This is a general feature, happening in all three models (tracking, scaling, and thawing).

Finally, using relation (\ref{Link_w0_p}), we deduce that our constraints on $w_{(0)}$ translate into $p<0.17$. This upper limit is less constraining than previously claimed in \cite{ChibaEtAl13}, even though we are using more recent data. This discrepancy is again simply because we did not fix the value of $H_0$.

\subsection{B. Scaling freezing models}

\begin{figure}
  \includegraphics[width=1.\linewidth]{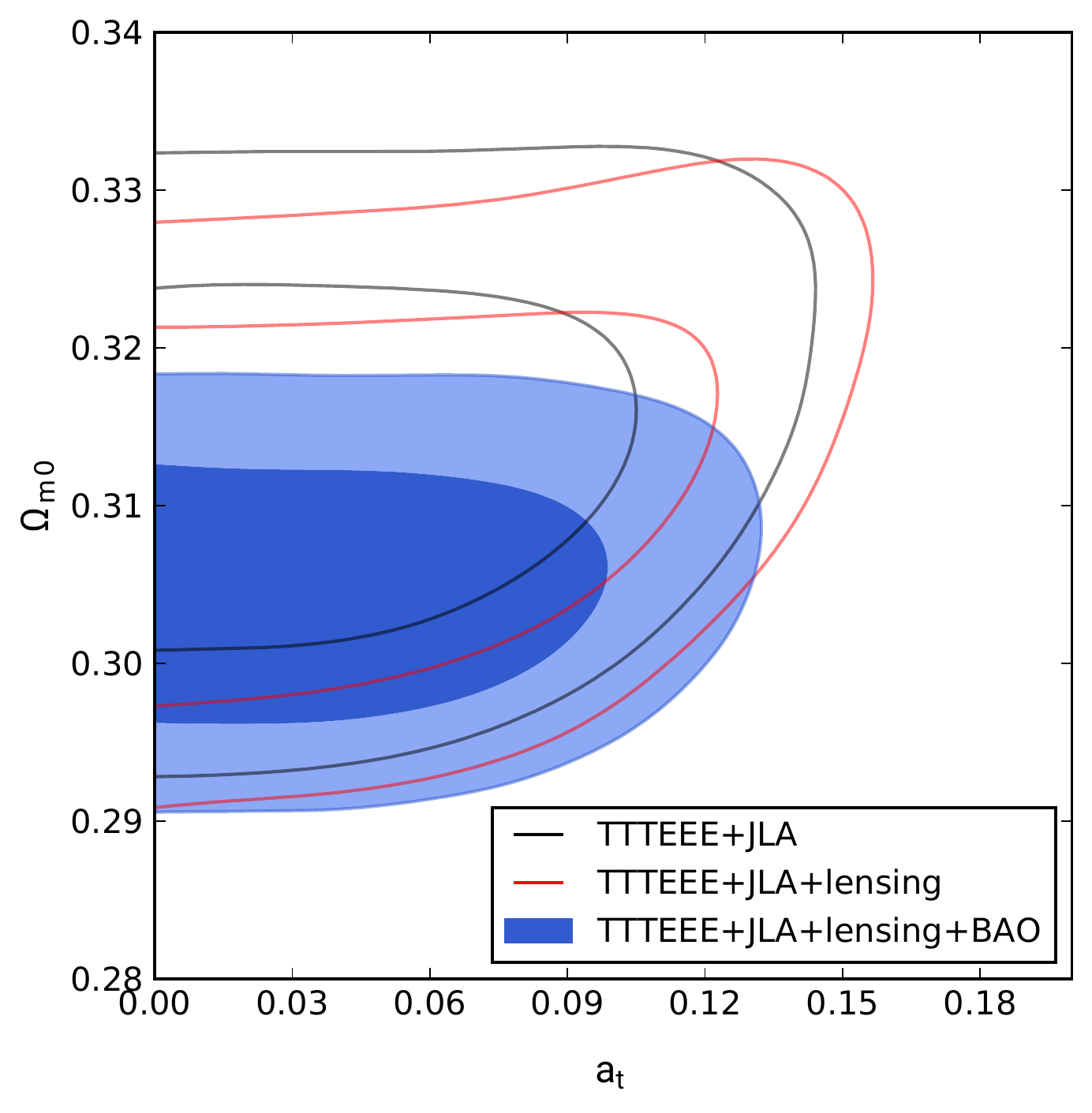}
  \caption{Observational constraints on models with scaling freezing equation of state parametrized by (\ref{EoS_Scaling}), in the $(a_t,\Omega_{m0})$ plane, having marginalized over all the other parameters. The red and black lines represent the constraints without BAO, with the inner lines corresponding to $1\sigma$ contours and the outer lines the $2\sigma$ ones. The filled contours in blue are the constraints including BAO measurements.}
  \label{fig:Contours_ScFr}
\end{figure}

\begin{figure}
  \includegraphics[width=1.1\linewidth]{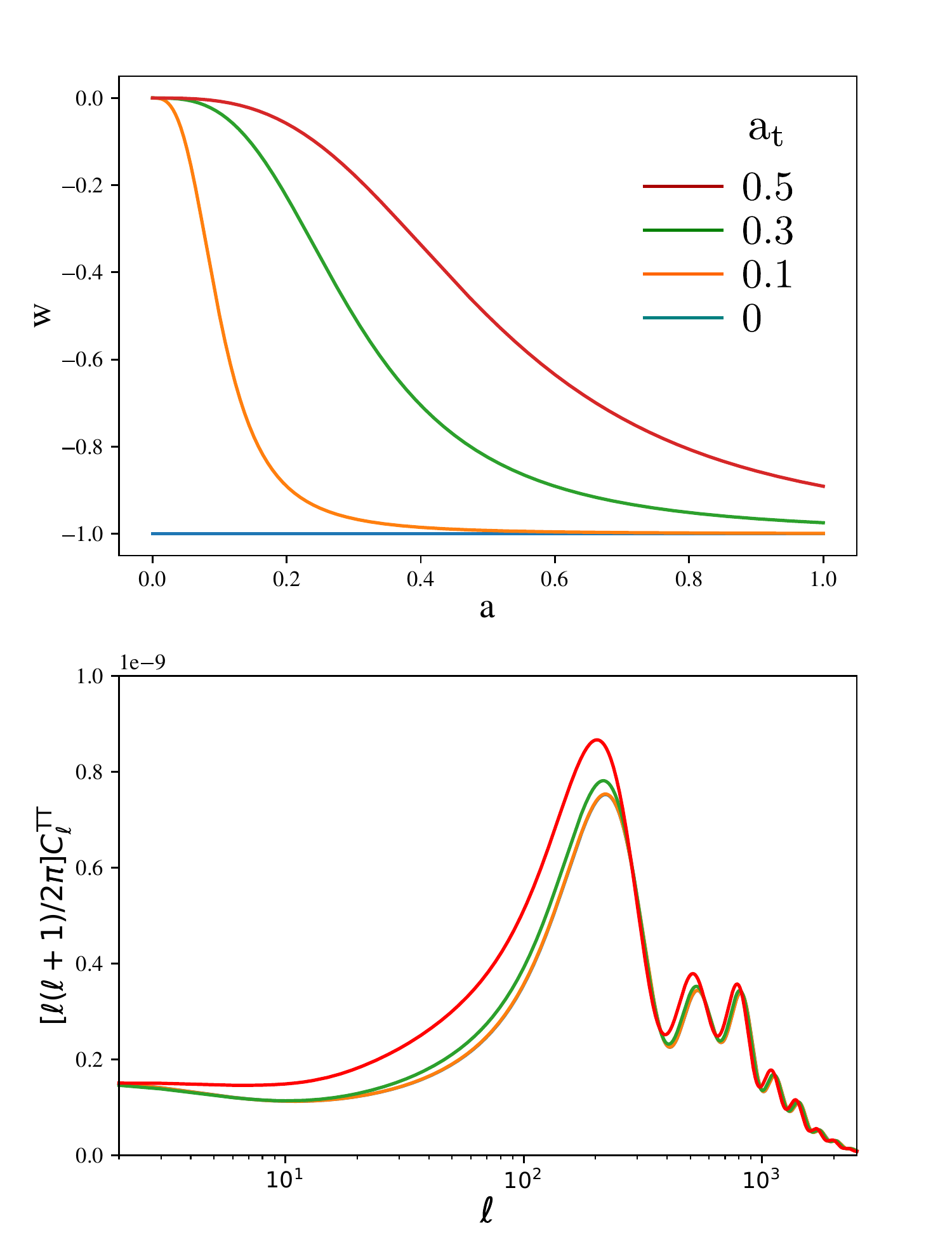}
  \caption{The bottom panel shows the power spectrum of the CMB temperature anisotropy for various scaling freezing models with equation of state shown in the top panel. The blue curve is the cosmological constant case. In the lower panel, the blue and orange lines overlap, i.e. for $a_t \lesssim 0.1$ the spectrum is only very weakly modified. For $a_t \gtrsim 0.1$ however, the effect becomes significant, so much that these models are strongly disfavored by the data as shown in fig~\ref{fig:Contours_ScFr}.}
  \label{fig:Interpretation_ScalingFreezing}
\end{figure}

Figure \ref{fig:Contours_ScFr} is our main result in the scaling freezing case. It corresponds to the $1\sigma$ and $2\sigma$ confidence regions in the $(a_t,\Omega_{m0})$ plane, where $\Omega_{m0}$ is the density parameter of matter today. Marginalizing over all the other parameters, we get the constraint
\begin{equation}
a_t < 0.11 \ \mathrm{i.e.} \ z_t > 8.1 \ (95 \% \ \mathrm{C.L.})
\end{equation}
for the transition scale factor. This result shows that the data favor a transition from a scaling matter epoch to the dark energy dominated era with equation of state close to $w = -1$ that occurs at a very early cosmological epoch, i.e. DE behaving as a cosmological constant. We interpret this as follows. For large $a_t$, scaling freezing quintessence behaves like matter ($w=0$) in the early universe, but without contributing to gravitational potentials because its sound speed is always equal to unity. Therefore it slows down the growth of structures and induces a large early ISW effect, which provides additional contribution to the CMB angular power spectrum on intermediate scales. To reinforce this assertion, we plot in Fig.~\ref{fig:Interpretation_ScalingFreezing} the power spectrum for various models, differing by their value of $a_t$. We see that indeed the larger $a_t$ is, the more the power spectrum is modified at intermediate multipoles $\ell$. Now, since JLA data probe low redshifts, the black line in Fig.~\ref{fig:Contours_ScFr} is essentially due to the CMB data. Therefore the fact that it indicates that $a_t \gtrsim 0.1$ models are incompatible with the measurements must be due to the aforementioned modification in the power spectrum which then becomes too strong.

\subsection{C. Thawing models}

\begin{figure}
  \includegraphics[width=1.\linewidth]{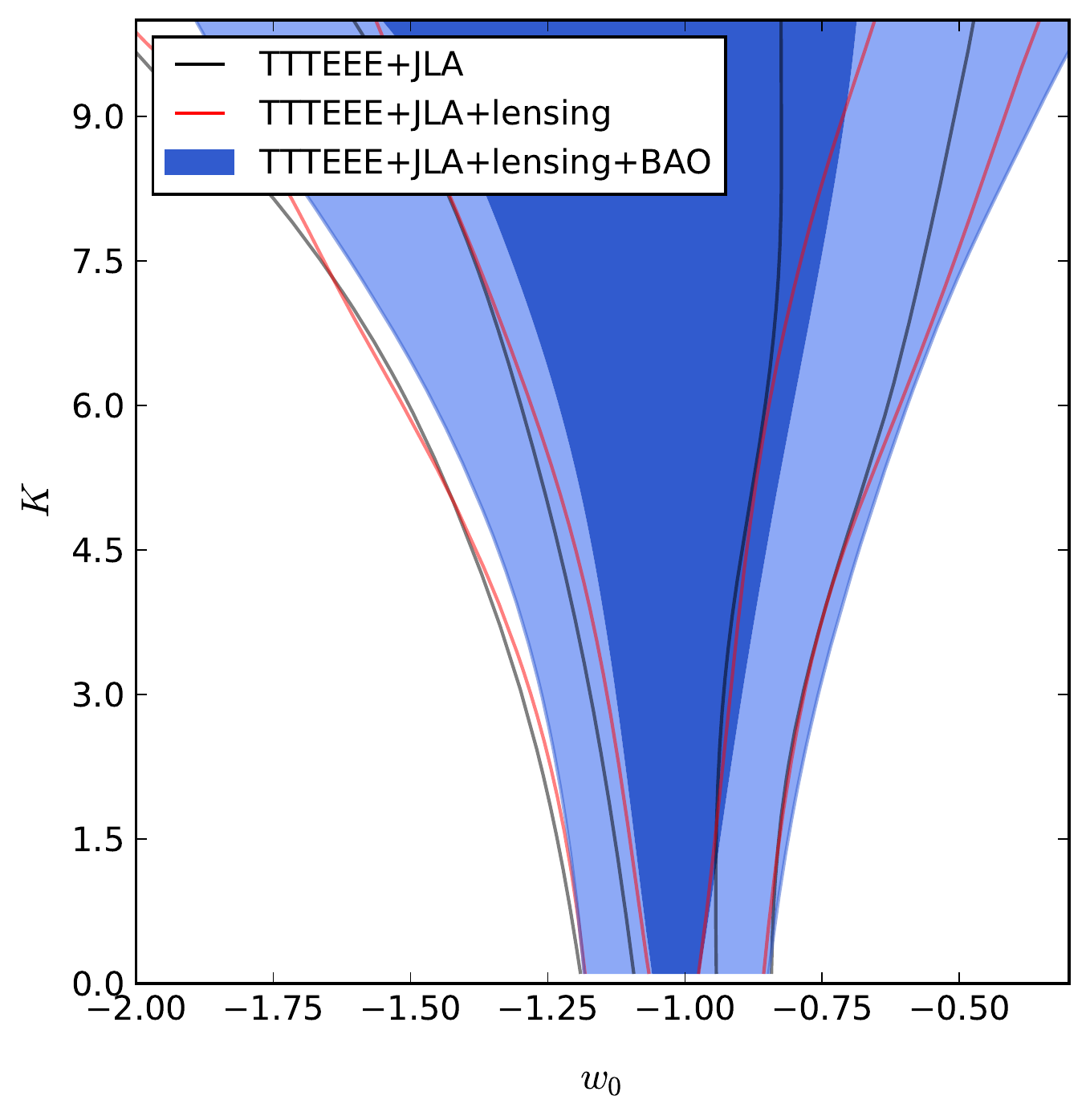}
  \caption{Observational constraints on models with thawing equation of state (\ref{EoS_Thawing}), in the $(w_0,K)$ plane, having marginalized over all the other parameters. The red and black lines represent the constraints without BAO, with the inner lines corresponding to $1\sigma$ contours and the outer lines the $2\sigma$ ones. The filled contours in blue are the constraints including BAO measurements. We set the prior $0.1 \leq K \leq 10$ for the analytic expression (\ref{EoS_Thawing}) to remain reliable, i.e. close to the numerical result.}
  \label{fig:Contours_Thaw_1}
\end{figure}

\begin{figure}
  \includegraphics[width=1.05\linewidth]{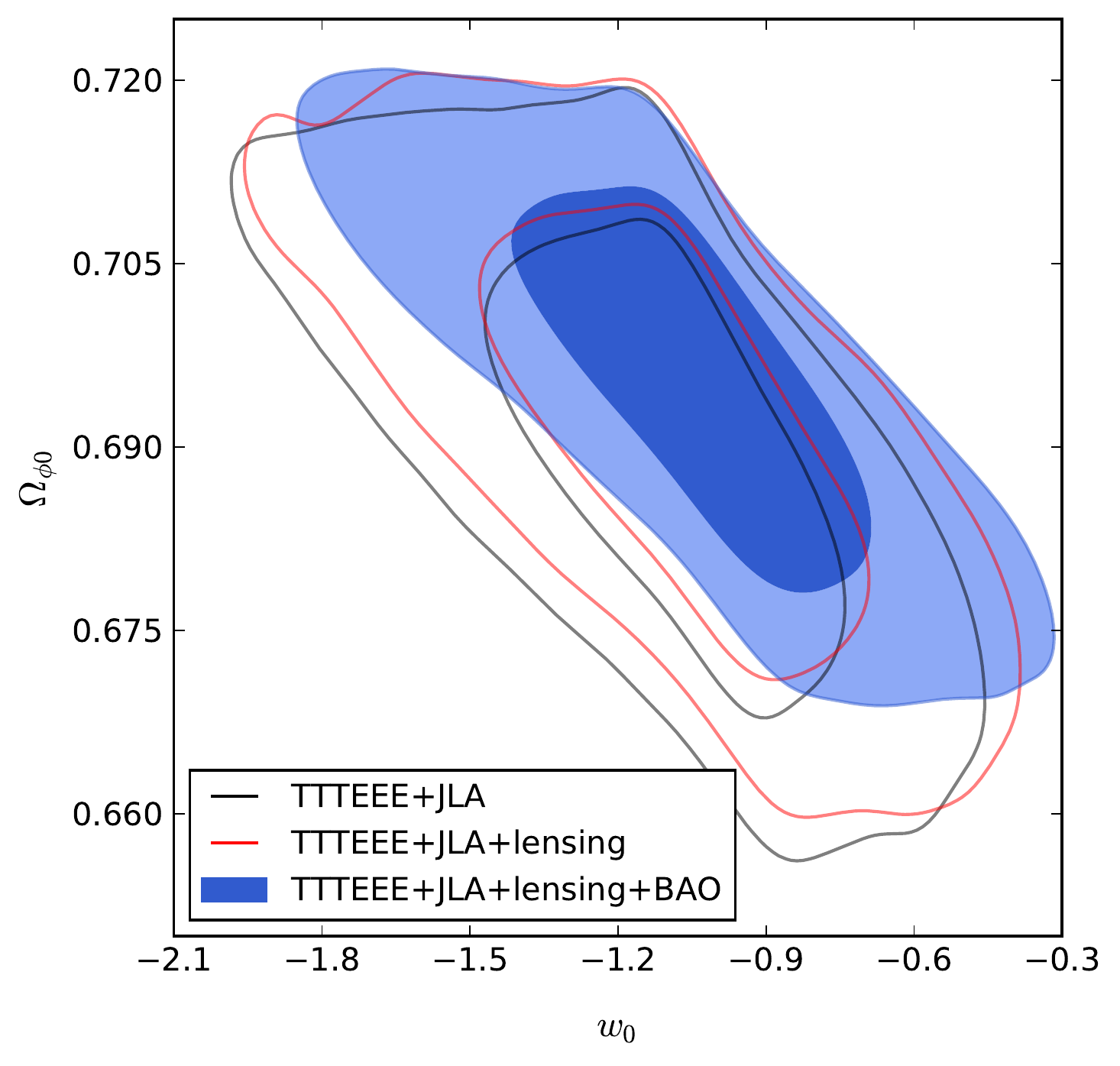}
  \caption{Observational constraints on models with thawing equation of state (\ref{EoS_Thawing}), in the $(w_0,\Omega_{\phi 0})$ plane, having marginalized over all the other parameters (with the prior $0.1 \leq K \leq 10$). The red and black lines represent the constraints without BAO, with the inner lines corresponding to $1\sigma$ contours and the outer lines the $2\sigma$ ones. The filled contours in blue are the constraints including BAO measurements.}
  \label{fig:Contours_Thaw_2}
\end{figure}

The observational constraints we obtain on the parameters $w_0$, $K$ and $\Omega_{\phi 0}$ intervening in the thawing case are presented in Figs.~\ref{fig:Contours_Thaw_1} and \ref{fig:Contours_Thaw_2}. They correspond, respectively, to the $1\sigma$ and $2\sigma$ confidence regions in the $(w_0,K)$ plane marginalizing over $\Omega_{\phi 0}$, and in the $(w_0,\Omega_{\phi 0})$ plane  marginalizing over $K$. They are plotted without the quintessence prior and in the same three cases (various combinations of data sets) as in Figs.~\ref{fig:Contours_Track} and \ref{fig:Contours_ScFr}. We observe that the BAO data now improve significantly the constraints only in the $(w_0,\Omega_{\phi 0})$ case. After marginalizing over $w_0$ we get, without the quintessence prior,
\begin{equation}
0.674 < \Omega_{\phi 0} < 0.717 \ (95 \% \ \mathrm{C.L.})
\end{equation}
while with the prior,
\begin{equation}
0.671 < \Omega_{\phi 0} < 0.703 \ (95 \% \ \mathrm{C.L.}).
\end{equation}
Now on the contrary, marginalizing over $\Omega_{\phi 0}$ we get, without the quintessence prior,
\begin{equation}
-1.70 < w_0 < -0.496 \ (95 \% \ \mathrm{C.L.})
\end{equation}
while with the prior,
\begin{equation}
-1 < w_0 < -0.473 \ (95 \% \ \mathrm{C.L.}).
\end{equation}

Compared to \cite{ChibaEtAl13}, generally speaking, our contours are shifted towards less negative values of $w_0$. This is once more essentially because these authors chose a value of $H_0$, and a relatively high one compared to the current Planck results. For this reason, our contraints tend to be more conservative. For example, without the quintessence prior and marginalizing over all the other parameters, in this case they obtained the upper limit $w_0 < -0.89 \ (95 \% \ \mathrm{C.L.})$ while we now have $w_0 < -0.496$ at the same confidence level. This is why our contours in Fig.~\ref{fig:Contours_Thaw_1} are not excluding $w_0 \lesssim -1$ as they stated, in particular with BAO data. On the contrary, however, they obtained as a lower limit $-2.18 < w_0 \ (95 \% \ \mathrm{C.L.})$ while we now have $-1.70 < w_0$ at the same confidence level. As far as the parameter $K$ is concerned, as can be seen in Fig.~\ref{fig:Contours_Thaw_1}, it is not constrained by the data, similarly as in the previous study of \cite{ChibaEtAl13}.

\subsection{D. Comparing the models}

The minimum effective $\chi^2$ for our three DE models are comparable to each other, namely 11756.4, 11755.8, and 11756.1, for the scaling, tracking, and thawing models, respectively, using the TT,TE,EE, lensing, JLA and BAO data. Now, the number of parameters in the two freezing cases are the same and the thawing one has only one additional parameter, so the significance, judged on Jeffreys' scale, is neither strong nor decisive \cite{Liddle07}. Therefore we conclude that none of these models is preferred from our analysis. In other words, the current observational data are not precise enough to distinguish between scaling, tracking, and thawing quintessence models \cite{2014JCAP...03..045T}.

\begin{figure}
  \includegraphics[width=1.\linewidth]{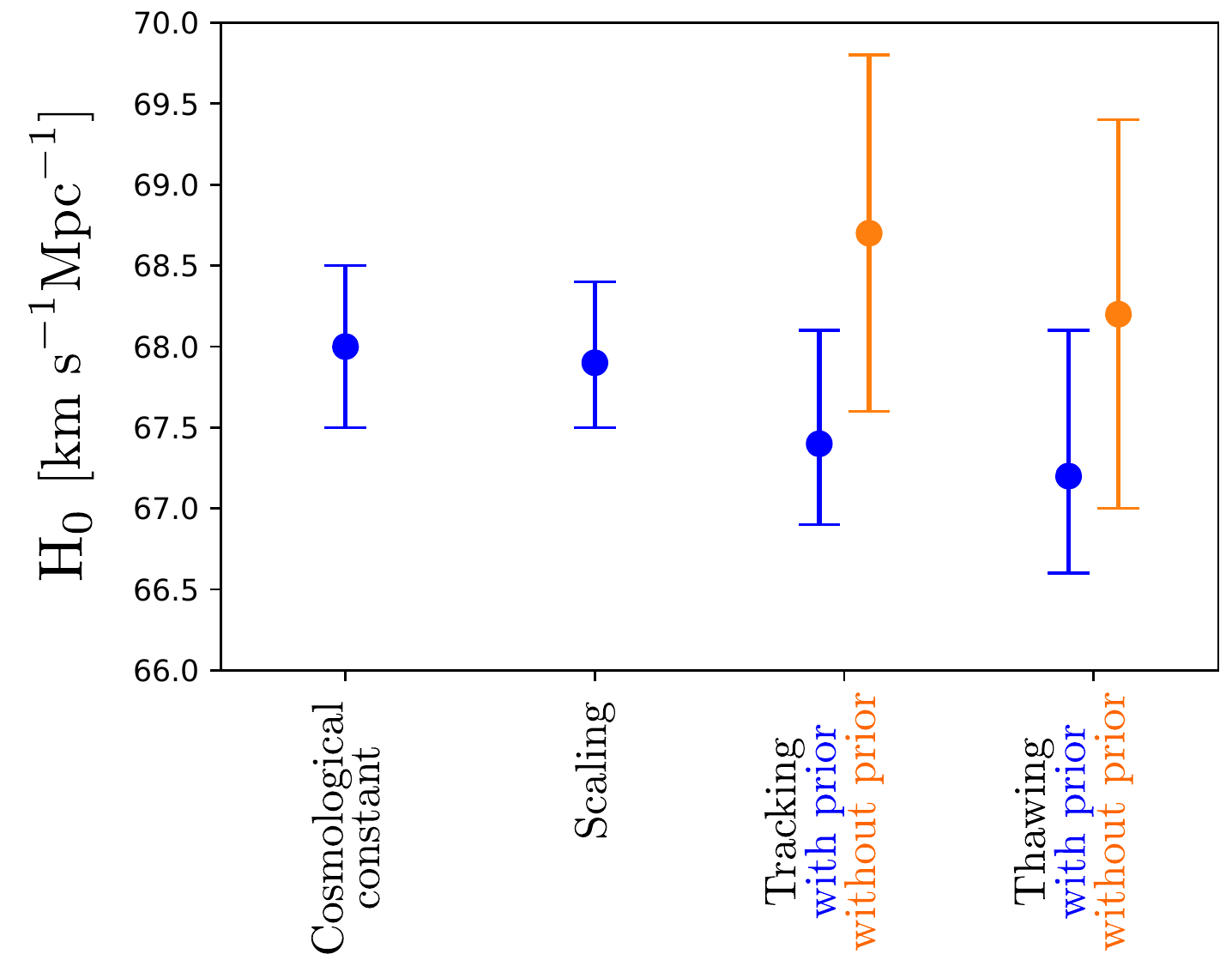}
  \caption{Constraints on the Hubble constant for the various DE models considered here, using the TT,TE,EE, lensing, JLA and BAO data. The dots correspond to the mean values of $H_0$, and the bars are the $1\sigma$ constraints. In blue the models correspond to quintessence, while we do not put the quintessence prior in the two orange cases.}
  \label{fig:H0}
\end{figure}

In Fig.~\ref{fig:H0} we show how including the various DE models affects the constraints on the Hubble constant. From this we conclude that while phantom DE models could in principle remedy the tension between the local measurement \cite{RiessEtAl16} and Planck results \cite{Planck16} mentioned in the introduction, such models are not preferred by the data, and it is even more unlikely in the case of quintessence.

\subsection{E. Considering massive neutrinos}

In the above, we neglected the mass of neutrinos. However, it is well known that they modify cosmic history, notably by washing out small scale structures. The properties of DE and that of neutrinos must, therefore, be correlated. In this section, we consider the presence of one massive neutrino of mass $M_\nu$, keeping the other two massless. In effect, the parameter $M_\nu$ then may be interpreted as the sum of neutrino masses, as if we had considered all three of them to be massive.

\begin{figure}
\includegraphics[scale=0.6]{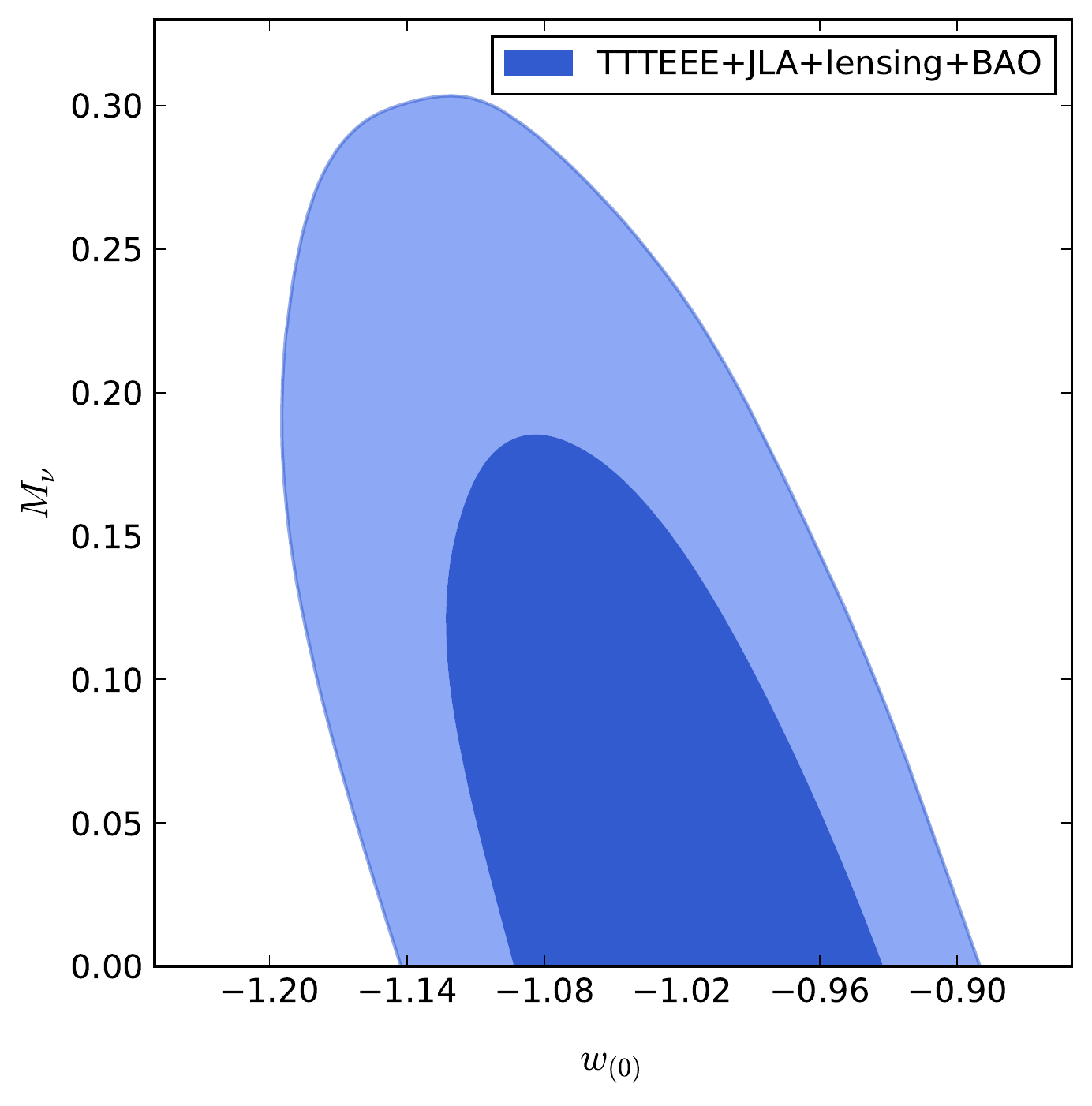}
\caption{$1\sigma$ (dark blue) and $2\sigma$ (light blue) observational contours (including BAO measurements) on models with tracking equation of state (\ref{EoS_Tracking}) with one massive neutrino of mass $M_\nu$ and two massless neutrinos, in the $(w_{(0)},M_\nu)$ plane having marginalized over all the other parameters.}
\label{fig:Neutrino_track}
\end{figure}

\begin{figure}
\includegraphics[scale=0.6]{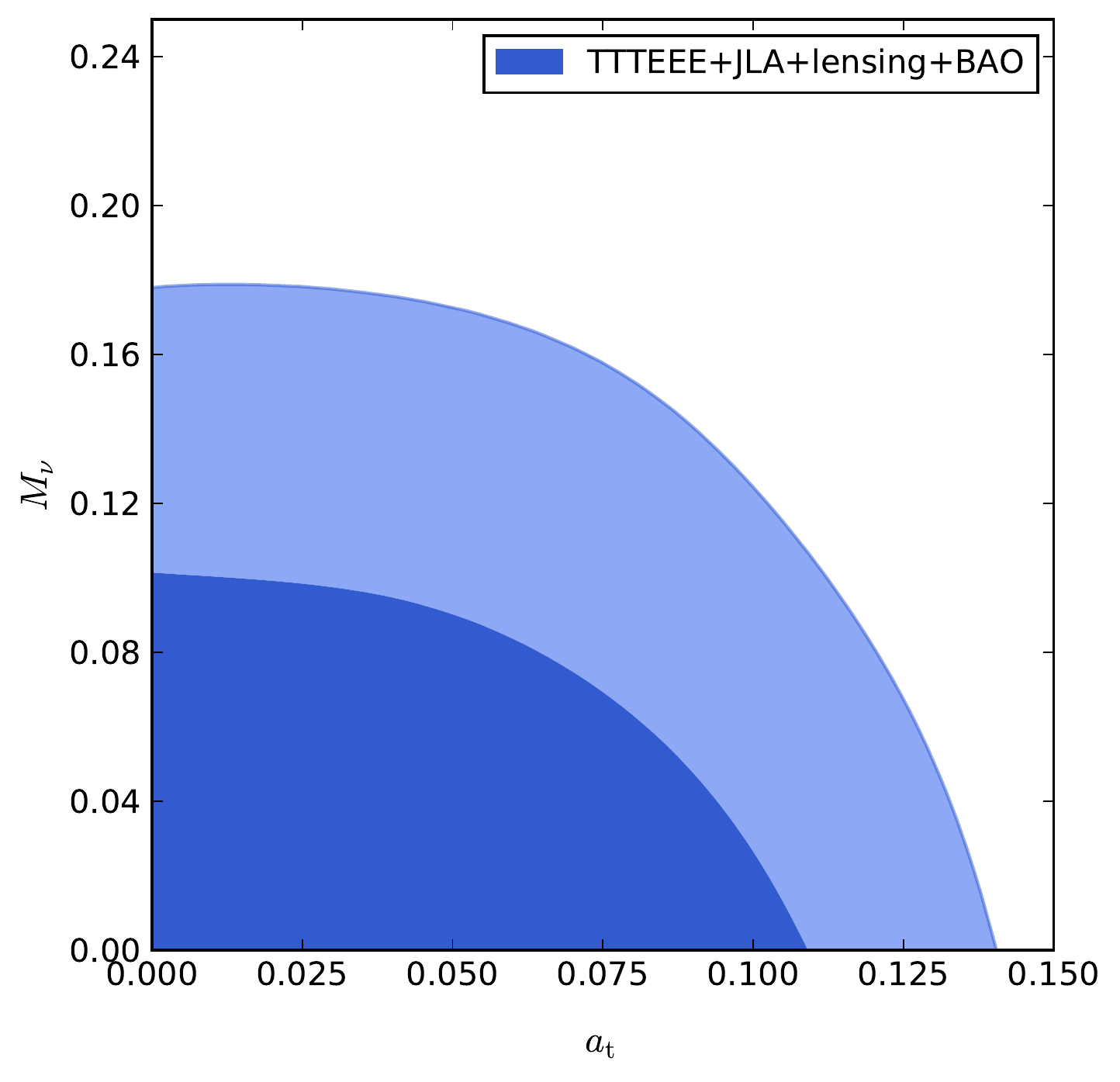}
\caption{$1\sigma$ (dark blue) and $2\sigma$ (light blue) observational contours (including BAO measurements) on models with scaling freezing equation of state parametrized by (\ref{EoS_Scaling}) with one massive neutrino of mass $M_\nu$ and two massless neutrinos, in the $(a_t,M_\nu)$ plane having marginalized over all the other parameters.}
\label{fig:Neutrino_scfr}
\end{figure}

\begin{figure}
\includegraphics[scale=0.6]{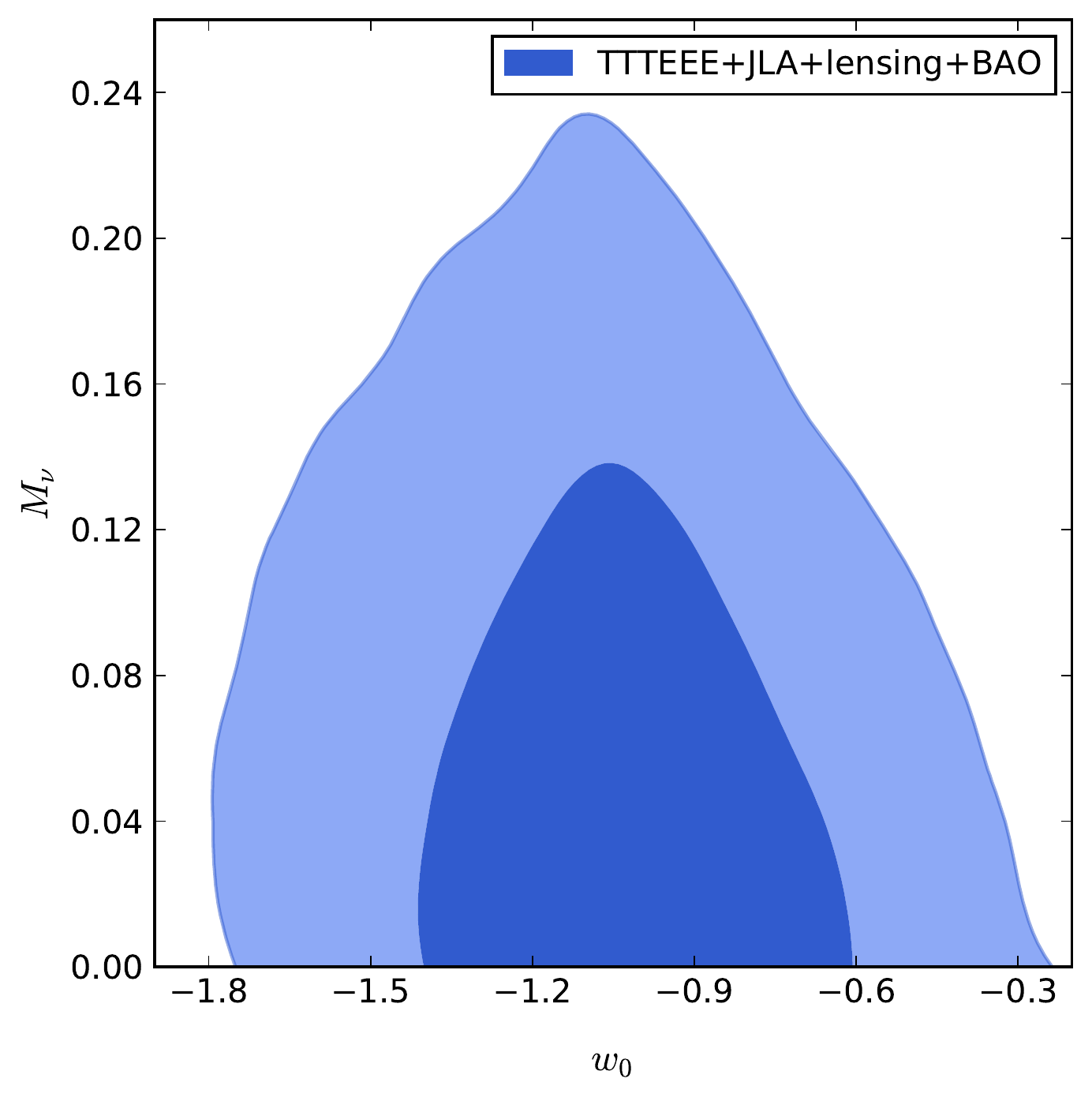}
\caption{$1\sigma$ (dark blue) and $2\sigma$ (light blue) observational contours (including BAO measurements) on models with thawing equation of state (\ref{EoS_Thawing}) with one massive neutrino of mass $M_\nu$ and two massless neutrinos, in the $(w_0,M_\nu)$ plane having marginalized over all the other parameters.}
\label{fig:Neutrino_thaw}
\end{figure}

In Figs.~\ref{fig:Neutrino_track}, \ref{fig:Neutrino_scfr}, and \ref{fig:Neutrino_thaw} we present the confidence regions we obtain in the $(w_{(0)},M_\nu)$, $(a_t,M_\nu)$ and $(w_0,M_\nu)$ planes for tracking, scaling, and thawing, respectively. These results are shown in the most constraining cases, namely with all our data (Planck temperature, polarization and lensing, JLA and BAO). The constraints on the total mass $M_\nu$ are as follows. In the tracking case, after marginalizing over all the remaining parameters, we get, without the quintessence prior,
\begin{equation}
M_\nu < 0.25 \ \mathrm{eV} \ (95 \% \ \mathrm{C.L.})
\end{equation}
while with the prior,
\begin{equation}
M_\nu < 0.15 \ \mathrm{eV} \ (95 \% \ \mathrm{C.L.}).
\end{equation}
In the scaling case, we get
\begin{equation}
M_\nu < 0.16 \ \mathrm{eV} \ (95 \% \ \mathrm{C.L.}).
\end{equation}
Finally, in the thawing case, without prior,
\begin{equation}
M_\nu < 0.17 \ \mathrm{eV} \ (95 \% \ \mathrm{C.L.})
\end{equation}
while with it,
\begin{equation}
M_\nu < 0.15 \ \mathrm{eV} \ (95 \% \ \mathrm{C.L.}).
\end{equation}

The tendencies in these figures can be qualitatively understood as follows. For tracking freezing models, larger $w_{(0)}$ will induce greater suppression in the growth of large scale structure. Therefore, large masses of neutrinos are not allowed for large $w_{(0)}$, hence, the degeneracy obtained in Fig.~\ref{fig:Neutrino_track}. For the same reason, for scaling freezing models, large masses of neutrinos are not compatible with a large transition scale factor $a_t$, because again a DE component with a large transition scale factor $a_t$ will cause a greater suppression of the large scale structure. In contrast, we do not find any clear degeneracy between $M_\nu$ and $w_0$ in thawing models. This might be due to the fact that in thawing models DE acts almost like the cosmological constant in the most part of the expansion history of the universe and the growth of structure is less affected by the parameter $w_0$, as long as $\Omega_\lambda$ is determined to give the right distances to the BAO and the last scattering surface.

\section{Conclusions}

We have updated the observational constraints on tracking freezing, scaling freezing, and thawing models for quintessence.

Compared to the previous study \cite{ChibaEtAl13} we let the value of $H_0$ vary, which is essential since its precise value is still under debate nowadays. Relaxing this assumption modifies significantly the constraints. For example the directions of degeneracy are in several cases totally different. But also, since we in addition used more recent data sets, our numerical values for the constraints differ significantly too. For instance our value on the $2\sigma$ upper limit on $w_{(0)}$ (and thus on $p$ for a power-law potential) of the tracking models is less constraining than previously claimed, while its lower limit however is improved.

The constraints on the freezing models are particularly stringent. In the tracking case with the quintessence prior, $-1 < w_{(0)} < -0.923 \ (95 \% \ \mathrm{C.L.})$; i.e., the equation of state has to be very close to $-1$. Similarly, in the scaling case studied here the transition to $w = -1$ occurs very early in the history of the Universe, with $a_t < 0.11 \ \mathrm{i.e.} \ z_t > 8.1 \ (95 \% \ \mathrm{C.L.})$.

Generally speaking, we observe the following trends: larger values of $H_0$ are accompanied by equations of state more negative; in most cases BAO data improve significantly the constraints; all in all, the data are consistent with the cosmological constant; and we conclude that such dynamical DE models are unlikely to remedy the tension between the local and CMB measurements.

Finally, we also considered the case of massive neutrinos and derived constraints of the sum of their mass in the various scenarios we considered.


\section{Acknowledgments}
We thank T.Chiba, S.Ili\'{c} and S.Tsujikawa for fruitful discussions, and the anonymous referee for his/her comments. This work is in part supported by MEXT Grant-in-Aid for Scientific Research on Innovative Areas No. 15H05890 (N.S. and K.I.), No. 15K2173 (N.S. and J.B.D.), No. 16H01543 (K.I.), and from JSPS [No. 16J05446 (J.O.)].



%

\end{document}